\documentclass[final,3p,times,twocolumn]{elsarticle}

 \usepackage{graphicx}

\usepackage{amssymb}

\journal{Solid State Communications}

\begin{document}

\begin{frontmatter}



\title{Iron-platinum-arsenide superconductors Ca$_{10}$(Pt$_n$As$_8$)(Fe$_{2-x}$Pt$_x$As$_2$)$_5$} 

\author[label1,label2]{Minoru Nohara\corref{cor1}}
\author[label1,label2]{Satomi Kakiya}
\author[label1,label2]{Kazutaka Kudo}
\author[label1]{Yoshihiro Oshiro}
\author[label1]{Shingo Araki}
\author[label1]{Tatsuo C. Kobayashi}
\author[label3]{Kenta Oku}
\author[label3]{Eiji Nishibori}
\author[label3]{Hiroshi Sawa}

\cortext[cor1]{Corresponding author. Tel.: +81-86-251-7828; fax: +81-86-251-7830. E-mail address: nohara@science.okayama-u.ac.jp}
\address[label1]{Department of Physics, Okayama University, Okayama 700-8530, Japan}
\address[label2]{JST, Transformative Research-Project on Iron Pnictides (TRIP), Tokyo 102-0075, Japan}
\address[label3]{Department of Applied Physics, Nagoya Univeristy, Nagoya 464-8603, Japan}

\begin{abstract}
An overview of the crystal structures and physical properties of the recently discovered iron-platinum-arsenide superconductors, 
Ca$_{10}$(Pt$_n$As$_8$)(Fe$_{2-x}$Pt$_x$As$_2$)$_5$ ($n$ = 3 and 4), which have a superconducting transition temperature up to 38 K, is provided. The crystal structure consists of superconducting Fe$_2$As$_2$ layers alternating with platinum-arsenic layers, Pt$_n$As$_8$. The upper critical field $H_{c2}$, hydrostatic pressure dependence of superconducting transition temperature $T_c$, and normal-state magnetic susceptibility are reported. 
\end{abstract}


\end{frontmatter}

\section{Introduction}
\label{}

The discovery of superconductivity at a superconducting transition temperature $T_c$ of 26 K in LaFeAsO$_{1-x}$F$_x$ has triggered an intensive exploration of novel iron-based superconductors \cite{doi:10.1021/ja800073m}. 
To date, a number of iron-based superconductors have been identified  \cite{JPSJ.78.062001}. 
Their crystal structure consists of alternately stacked two-dimensional Fe$_2$As$_2$ layers, in which high-$T_c$ superconductivity emerges, and spacer layers. These superconducting materials can be classified into three groups in terms of spacer layers, as visually summarized in Fig.~1.
The first group of materials consists of LiFeAs \cite{PhysRevB.78.060505} and BaFe$_2$As$_2$ \cite{PhysRevLett.101.107006}, in which the spacer layers comprise alkali ions or alkaline-earth ions. 
The second group consists of LaFeAsO \cite{doi:10.1021/ja800073m} and CaFeAsF  \cite{JACS_130_14428_2008}, in which the spacer layers are composed of slabs of rare-earth oxides or alkaline-earth fluorides with a fluorite-type structure. 
The third group consists of the materials in which the spacer layers are composed of complex metal oxides i.e., Sr$_3$Sc$_2$O$_5$Fe$_2$As$_2$ \cite{PhysRevB.79.024516} and Sr$_4$(Sc,Ti)$_3$O$_8$Fe$_2$As$_2$ \cite{APEX.3.063102} with perovskite-type spacer layers
and Sr$_4$V$_2$O$_6$Fe$_2$As$_2$ \cite{PhysRevB.79.220512} and Ca$_4$(Al,Ti)$_2$O$_6$Fe$_2$As$_2$ \cite{SST-23-11-115005, shirage:172506} with a combination of perovskite-type and rocksalt-type spacer layers and 
their homologous series compounds  \cite{APEX.3.063102, SST-22-8-085001, APEX.2.063007, SST-23-4-045001, ogino:072506}.

\begin{figure*}[t]
\begin{center}
\includegraphics[width=16cm]{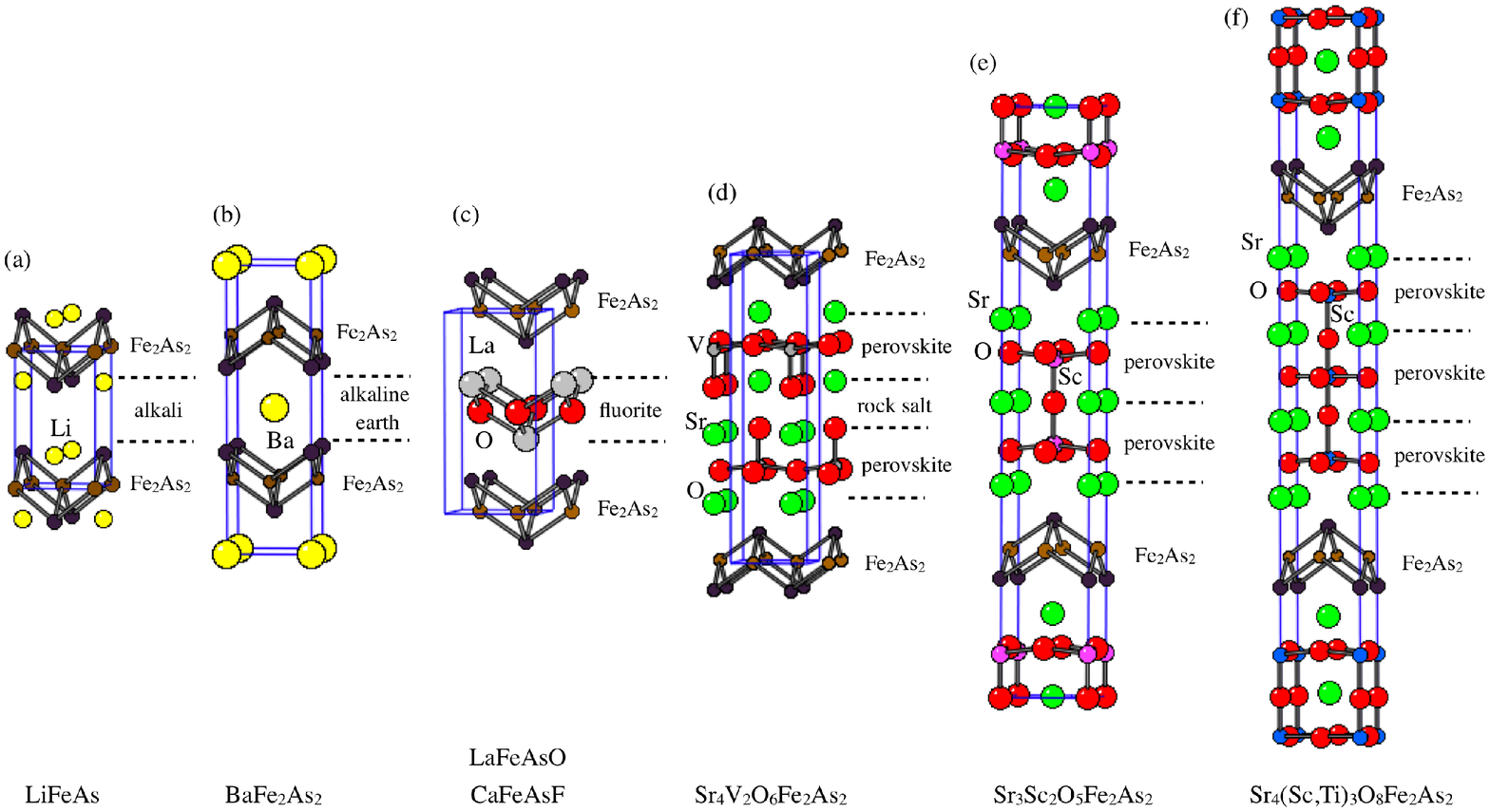}
\caption{
\label{}
Crystal structures of various iron-based superconductors categorized by the structure type of spacer layers: (a) LiFeAs, (b) BaFe$_2$As$_2$, (c) LaFeAsO and CaFeAsF, (d) Sr$_4$V$_2$O$_6$Fe$_2$As$_2$, (e) Sr$_3$Sc$_2$O$_5$Fe$_2$As$_2$, and (f) Sr$_4$(Sc,Ti)$_3$O$_8$Fe$_2$As$_2$.
}
\end{center}
\end{figure*}

All of these spacer layers are electrically inert because of their strong ionic chemical bonds; thus, no atomic orbitals of the spacer layers mix with the Fe $3d$ orbitals of the superconducting Fe$_2$As$_2$ layers. Consequently the primary role of the spacer layers is to separate the Fe$_2$As$_2$ layers in order to realize nearly two-dimensional electronic band structures originating from the square lattice of iron (Fe). The parent materials exhibit antiferromagnetic (AFM) ordering at low temperatures because of the characteristic nesting between the hole Fermi surfaces centered at the $\Gamma$ point and the electron Fermi surfaces centered at  the M point \cite{JPSJ.78.062001}. 
The spacer layers, in a secondary role, supply charge carriers to the Fe$_2$As$_2$ layers to suppress the AFM ordering and drive the system into a superconducting state. This can be best achieved by partial chemical substitutions for the constituent elements of the spacer layers. For instance, LaFeAsO$_{1-x}$F$_x$ exhibits superconductivity at 26 K by partial substitution of F$^{-}$ for O$^{2-}$ to introduce electron carriers \cite{doi:10.1021/ja800073m}. Ba$_{1-x}$K$_x$Fe$_2$As$_2$ exhibits superconductivity at 38 K by partial chemical substitutions of K$^{+}$ for Ba$^{2+}$ to introduce hole carriers \cite{PhysRevLett.101.107006}. 
Third, the spacer layers act to tune the superconducting Fe$_2$As$_2$ layers to optimize superconductivity. The most remarkable example is the increase in $T_c$ by replacing La with smaller rare-earth ions. The highest $T_c$ value of 56 K is achieved in Th-substituted GdFeAsO \cite{0295-5075-83-6-67006}. The replacement of La by the smaller Gd in the spacer layers leads to the modification of the bond angle; superconductivity is thought to be optimized when the As-Fe-As bond angle is close to that of the regular tetrahedron \cite{JPSJ.77.083704}.
Thus, the central issues for realizing higher $T_c$ are to find novel spacer layers and to engineer them to tune Fe$_2$As$_2$ layers, which exhibit high-$T_c$ superconductivity.

Recently, we discovered novel iron-based superconductors Ca$_{10}$(Pt$_n$As$_8$)(Fe$_{2-x}$Pt$_x$As$_2$)$_5$ with $n$ = 4 (referred to as $\alpha$-phase) and $n$ = 3 (referred to as $\beta$-phase) \cite{JPSJ.80.093704}.
These materials can be characterized by their unique spacer layers, namely platinum-arsenide $Pt_n$As$_8$, which has not been observed in previous iron-based superconductors. 
Both compounds crystallize in triclinic structures (space group $P\bar{1}$), in which Fe$_2$As$_2$ layers alternate with Pt$_n$As$_8$ spacer layers, as shown in Fig.~2. Superconductivity with a transition temperature of up to $T_c$ = 38 K is observed in the $\alpha$-phase ($n$ = 4), while the $\beta$-phase ($n$ = 3) exhibits superconductivity at 13 K \cite{JPSJ.80.093704}.

Two other groups have almost simultaneously reported similar results, motivated by our previous results \cite{nohara}. 
Ni {\it et al.} identified two phases Ca$_{10}$(Pt$_4$As$_8$)(Fe$_{2}$As$_2$)$_5$ (space group $P4/n$) and Ca$_{10}$(Pt$_3$As$_8$)(Fe$_{2-x}$Pt$_x$As$_2$)$_5$ (space group $P\bar{1}$) with $T_c$ = 25 K and 11 K, respectively \cite{Ni.arXiv}.
L{$\rm \ddot{o}$}hnert {\it et al.} identified three phases (CaFe$_{1-x}$Pt$_x$As)$_{10}$Pt$_3$As$_8$ (space group $P\bar{1}$), $\alpha$-(CaFe$_{1-x}$Pt$_x$As)$_{10}$Pt$_{4-y}$As$_8$ (space group $P4/n$), and $\beta$-(CaFe$_{1-x}$Pt$_x$As)$_{10}$Pt$_{4-y}$As$_8$ (space group $P\bar{1}$), in which $T_c$ of up to 35 K was observed \cite{Lohnert.arXiv}. 
Details of the crystal structures are different; the relation between the chemical composition and the superconducting properties of the materials has not been completely determined thus far. It is noteworthy that superconductivity is present in the $n$ = 4 member of Ca$_{10}$(Pt$_n$As$_8$)(Fe$_{2-x}$Pt$_x$As$_2$)$_5$ at up to 38 K.

The objective of this study is twofold. First, we provide an overview of the crystal structure of the novel iron-platinum arsenide superconductors Ca$_{10}$(Pt$_n$As$_8$)(Fe$_{2-x}$Pt$_x$As$_2$)$_5$.
Secondly, we present the latest experimental results including the upper critical field $H_{c2}$, hydrostatic-pressure effect on $T_c$, and magnetic susceptibility in the normal state; these results have not been provided in our previous report \cite{JPSJ.80.093704}.

\begin{figure*}[t]
\begin{center}
\includegraphics[width=9cm]{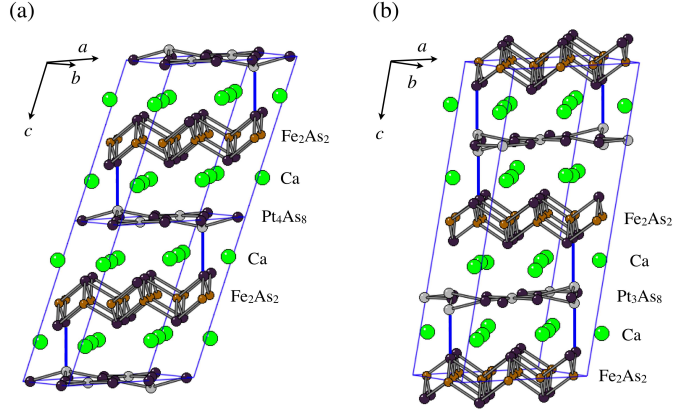}
\caption{
\label{}
Crystal structures of (a) Ca$_{10}$(Pt$_4$As$_8$)(Fe$_{2-x}$Pt$_x$As$_2$)$_5$ ($\alpha$-phase) and (b) Ca$_{10}$(Pt$_3$As$_8$)(Fe$_{2-x}$Pt$_x$As$_2$)$_5$ ($\beta$-phase) \cite{JPSJ.80.093704}. Thin solid lines represent unit cells. Two unit cells are shown for the $\alpha$-phase along the $c$ axis, while one unit cell is shown for the $\beta$-phase. Thick solid lines represent Pt-As bonds between the Fe$_2$As$_2$ layers and Pt$_n$As$_8$ layers.
}
\end{center}
\end{figure*}

\begin{figure}[t]
\begin{center}
\includegraphics[width=6cm]{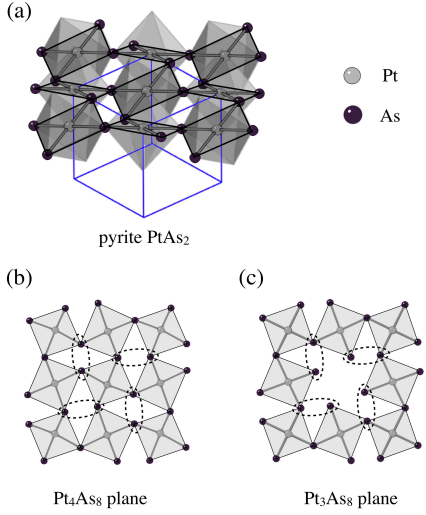}
\caption{
\label{}
Crystal structures of Pt$_n$As$_8$ layers. (a) Crystal structure of PtAs$_2$ showing a cubic pyrite-type structure (space group $Pa\bar{3}$). The structure consists of  a three-dimensional network of corner-sharing PtAs$_6$ octahedra that form As$_2$ dimers. A slab of Pt$_4$As$_8$  layer can be derived from the $ab$-plane of the pyrite structure. (b) Details of the Pt$_4$As$_8$ layer. (c) Details of the Pt$_3$As$_8$ layer. The dashed ellipsoids represent As$_2$ dimers.
}
\end{center}
\end{figure}

\section{Experimental}

Single crystals of Ca$_{10}$(Pt$_n$As$_8$)(Fe$_{2-x}$Pt$_x$As$_2$)$_5$ were grown as described in Ref. \cite{JPSJ.80.093704}.
Crystals as large as 1 $\times$ 1 $\times$ 0.1 mm$^3$ were obtained. 
Details of the structural analysis are provided in Ref. \cite{JPSJ.80.093704}. 
Resistivity in a magnetic field was measured using a Physical Property Measurement System (Quantum Design). Magnetization of powder samples was measured using a SQUID magnetometer (Quantum Design). 
Resistivity measurements under hydrostatic pressure were performed using an indenter cell \cite{10.1063/1.2459512}.

\section{Crystal Structure}

We identified two structural phases as depicted in Fig.~2: Ca$_{10}$(Pt$_4$As$_8$)(Fe$_{2-x}$Pt$_x$As$_2$)$_5$ ($\alpha$-phase) and Ca$_{10}$(Pt$_3$As$_8$)(Fe$_{2-x}$Pt$_x$As$_2$)$_5$ ($\beta$-phase) \cite{JPSJ.80.093704}. 
The corresponding crystallographic data are summarized in Table 1.
Both phases crystallize in triclinic structures (space group $P\bar{1}$).
The structures consist of alternately stacked (Fe$_2$As$_2$)$_5$ and Pt$_n$As$_8$ layers ($n$ = 4 for the $\alpha$-phase and $n$ = 3 for the $\beta$-phase) with five Ca ions between them.
The platinum-arsenide layers are characterized by a distorted square lattice of corner-sharing PtAs$_4$ squares, as shown in Fig.~3. 
Rotations of the PtAs$_4$ squares result in the formation of As$_2$ dimers. Such As$_2$ dimers are observed in PtAs$_2$ with a cubic pyrite-type structure (space group $Pa\bar{3}$); the Pt$_4$As$_8$ layers can be derived from the slab of the $ab$-plane of pyrite PtAs$_2$, as shown in Fig.~3(a). 

The formal electron counts of As$_2$ dimers and isolated As are [As$_2$]$^{4-}$ and As$^{3-}$, respectively. All the As atoms form dimers in the Pt$_n$As$_8$ layers and all the As atoms are isolated in the Fe$_2$As$_2$ layers. Thus, according to the charge balance, we estimate a formal electron count to be Fe$^{2+}$ and Pt$^{2+}$ for the $\beta$-phase, Ca$_{10}$(Pt$_3$As$_8$)(Fe$_{2-x}$Pt$_x$As$_2$)$_5$, when $x$ = 0.0. Thus, Ca$_{10}$(Pt$_3$As$_8$)(Fe$_{2}$As$_2$)$_5$ can be viewed as the parent compound. 
Thus far, compounds with $x$ = 0.0 have not been obtained; however, Ca$_{10}$(Pt$_3$As$_8$)(Fe$_{2-x}$Pt$_x$As$_2$)$_5$ has a Pt content of $x$ $\simeq$ 0.16. 
Partial substitution of Pt for Fe in the Fe$_2$As$_2$ layers leads to electron doping, thereby causing an under-doped regime in the $\beta$-phase with $x$ $\simeq$ 0.16, as confirmed by Hall measurements \cite{JPSJ.80.093704}.
Further electron doping is realized for the $\alpha$-phase, Ca$_{10}$(Pt$_4$As$_8$)(Fe$_{2-x}$Pt$_x$As$_2$)$_5$, owing to an increase in the Pt content in the Pt$_4$As$_8$ layers together with an increase in the Pt content $x$ ($\simeq$ 0.36) in the Fe$_2$As$_2$ layers, as indicated by Hall measurements \cite{JPSJ.80.093704}. 

The size of the Pt square lattice (with a Pt-Pt distance of approximately 4.4 \AA) is by far larger than the size of the Fe$_2$As$_2$  square lattice (approximately 3.9 {\AA} for CaFe$_2$As$_2$). This lattice mismatch leads to a structural distortion in the Fe$_2$As$_2$ layers so that the As-Fe-As bond angle approaches the ideal value, 109.47$^\circ$.
For the $\alpha$-phase, in which superconductivity was observed at temperatures up to 38 K, the As-Fe-As bond angle $\alpha$ lies between 109.08$^\circ$ and 109.55$^\circ$, depending on the five Fe sites. 
In contrast, for the $\beta$-phase with lower $T_c$, the FeAs$_4$ tetrahedra are distorted from the regular tetrahedron structure;  As-Fe-As bond angle $\alpha$ lies between 106.92$^\circ$ and 110.09$^\circ$, depending on the ten Fe sites.
This observation is in accordance with the fact that the maximum value of $T_c$ is higher when the bond angle of As-Fe-As is closer to the ideal value of 109.47$^\circ$ \cite{JPSJ.77.083704}.  

The Pt$_n$As$_8$ layers are not flat; however part of Pt ions are located at off-centered sites. A Pt ion that is located at such a site forms a chemical bond with the As ion at the adjacent Fe$_2$As$_2$ layers, as indicated by the thick solid lines in Figs. 2(a) and 2(b). The Pt-As bonds perpendicular to the $ab$-plane are reminiscent of the chemical bonds of SrPt$_2$As$_2$ with a CaBe$_2$Ge$_2$-type structure, in which a three-dimensional Pt-As network is formed \cite{JPSJ.79.123710}. Interestingly, SrPt$_2$As$_2$ exhibits superconductivity at 5.2 K \cite{JPSJ.79.123710}. Thus, it is speculated that the Pt$_n$As$_8$ layers are conducting and that the Pt $5d$ orbital may contribute to superconductivity in the present compounds. Band calculations suggest small but finite Pt contributions to the density of state at the Fermi level \cite{Lohnert.arXiv,Shein.arXiv}.

\begin{table}
\caption{
Crystallographic data of Ca$_{10}$(Pt$_4$As$_8$)(Fe$_{2-x}$Pt$_x$As$_2$)$_5$ with $x$ $\simeq$ 0.36  ($\alpha$-phase) and Ca$_{10}$(Pt$_3$As$_8$)(Fe$_{2-x}$Pt$_x$As$_2$)$_5$ with $x$ $\simeq$ 0.16 ($\beta$-phase) \cite{JPSJ.80.093704}.
}
\begin{tabular}{l@{\hspace{1.7cm}}c@{\hspace{1cm}}c}
\hline
label & $\alpha$-phase & $\beta$-phase\\
space group& $P\bar{1}$ & $P\bar{1}$ \\
$a$ (\AA) & 8.719(1) & 8.795(3)  \\
$b$ (\AA) & 8.727(1) & 8.789(3)  \\
$c$ (\AA) & 11.161(1)  & 21.008(7) \\
$\alpha$ ($^\circ$) & 99.04(2) &  94.82(8) \\
$\beta$  ($^\circ$) & 108.21(2) & 99.62(9) \\
$\gamma$  ($^\circ$) & 90.0(2) & 89.99(3) \\
Pt content $x$ & 0.36(4) &  0.16(1)\\
\hline
\end{tabular}
\end{table}

\section{Upper Critical Field}

\begin{figure}[t]
\begin{center}
\includegraphics[width=7cm]{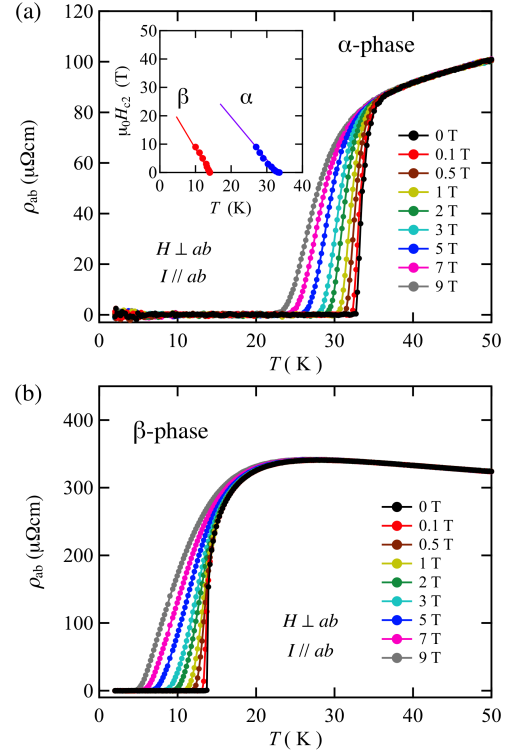}
\caption{
\label{}
Temperature dependence of in-plane resistivity $\rho_{ab}$ in magnetic fields $H$ $\perp$ $ab$ up to 9 T for (a) $\alpha$-phase and for (b) $\beta$-phase of Ca$_{10}$(Pt$_n$As$_8$)(Fe$_{2-x}$Pt$_x$As$_2$)$_5$. The inset shows the temperature dependence of the upper critical field $H_{c2}$ perpendicular to the $ab$-plane.
}
\end{center}
\end{figure}

Electrical resistivity of Ca$_{10}$(Pt$_4$As$_8$)(Fe$_{2-x}$Pt$_x$As$_2$)$_5$ ($\alpha$-phase) and Ca$_{10}$(Pt$_3$As$_8$)(Fe$_{2-x}$Pt$_x$As$_2$)$_5$ ($\beta$-phase) is shown in Figs.~4(a) and 4(b), respectively. The resistivity of the $\alpha$-phase exhibits metallic behavior over a wide temperature range \cite{JPSJ.80.093704}. The resistivity starts to decrease at approximately 37 K. The 10$-$90\% transition width is approximately 1.7 K, and the onset temperature determined from the 10\% rule is 34.6 K. Zero resistivity is observed at 32.7 K. 
In contrast, the resistivity of the $\beta$-phase shows semiconducting behavior below approximately 110 K \cite{JPSJ.80.093704}. The resistive transition is considerably broad. Zero resistivity is observed at 13.7 K. 

Figure 4 also shows the temperature dependence of in-plane resistivity $\rho_{ab}$ at various magnetic fields applied perpendicular to the $ab$-plane. With increasing field, $T_c$ decreases and the transition width is broadened.
The inset of Fig.~4(a) shows the plot the upper critical field $H_{c2}$ perpendicular to the $ab$-plane determined by the midpoint of the resistive transition as a function of temperature. The slopes of $H_{c2}$ at $T_c$ are $- 1.6$ T/K and $- 2.3$ T/K for the $\alpha$- and $\beta$-phases, respectively. 
From the Werthamer-Helfand-Hohenberg theory \cite{PhysRev.147.295}, which describes the orbital depairing field of conventional dirty type-II superconductors, we estimate the values of $H_{c2}(0)$ $=$ $-0.69T_c dH_{c2}/dT|_{T = T_c}$ $\sim$ 35 T and $\sim$ 22 T for the $\alpha$- and $\beta$-phases, respectively. These values are comparable to those of electron-doped Ba(Fe$_{1-x}$Co$_x$)$_2$As$_2$ \cite{JPSJ.78.023702}.  
In contrast, the transition width in magnetic field is broader in the present compounds, suggesting that the electronic states are more two dimensional in the present compounds than that in Ba(Fe$_{1-x}$Co$_x$)$_2$As$_2$.

\section{Effects of Hydrostatic Pressure on $T_c$}

\begin{figure}[t]
\begin{center}
\includegraphics[width=7cm]{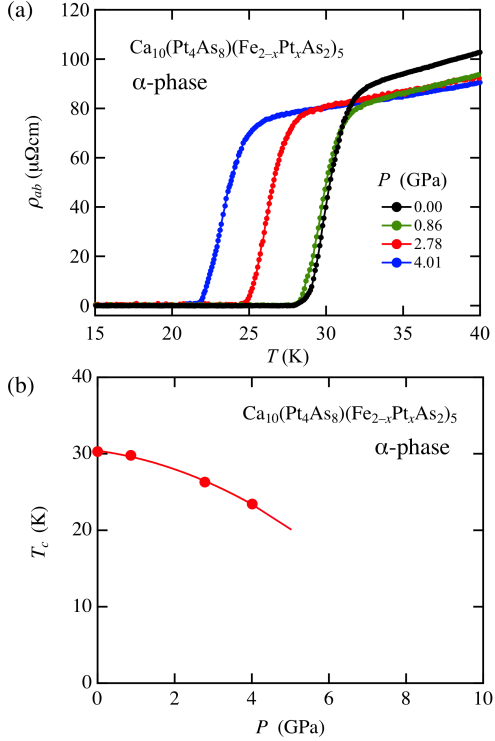}
\caption{
\label{}
(a) Temperature dependence of in-plane resistivity $\rho_{ab}$ at hydrostatic pressures up to 4.01 GPa for $\alpha$-phase of Ca$_{10}$(Pt$_n$As$_8$)(Fe$_{2-x}$Pt$_x$As$_2$)$_5$. (b) Superconducting transition temperature $T_c$ under high pressure. The solid line is guide for eyes.
}
\end{center}
\end{figure}

The temperature dependence of the in-plane electrical resistivity of the $\alpha$-phase Ca$_{10}$(Pt$_4$As$_8$)(Fe$_{2-x}$Pt$_x$As$_2$)$_5$ for pressures up to 4 GPa is shown in Fig. 5(a). This specimen exhibits a lower $T_c$ value than that used for the resistivity measurements in magnetic fields (in Fig.~4). No broadening of the transitions is observed with increasing pressure, thereby implying that sample inhomogeneities are sufficiently small. The pressure dependence of $T_c$ obtained from the midpoint of resistive transition is shown in Fig. 5(b). The $T_c$ value decreases with an initial slope of approximately $-$ 0.9 K/GPa. $T_c$ decreases rapidly at higher pressures.

The small initial slope $dT_c/dP$ may indicate that the specimen used is not optimally doped; $T_c$ may increase by further doping. 
Indeed, for SmFeAsO$_{1-x}$F$_x$,  the pressure coefficient $dT_c/dP$ is approximately $+$ 2.6 K/GPa at $x$ = 0.10 with $T_c$ = 17 K, while it is approximately $-$ 1.44 K/GPa at $x$ = 0.20 with $T_c$ = 49 K; a small coefficient is observed at $x$ = 0.15 with $T_c$ = 40 K \cite{doi:10.1021/ja8036838}.

\section{Magnetic Properties}

\begin{figure}[t]
\begin{center}
\includegraphics[width=7cm]{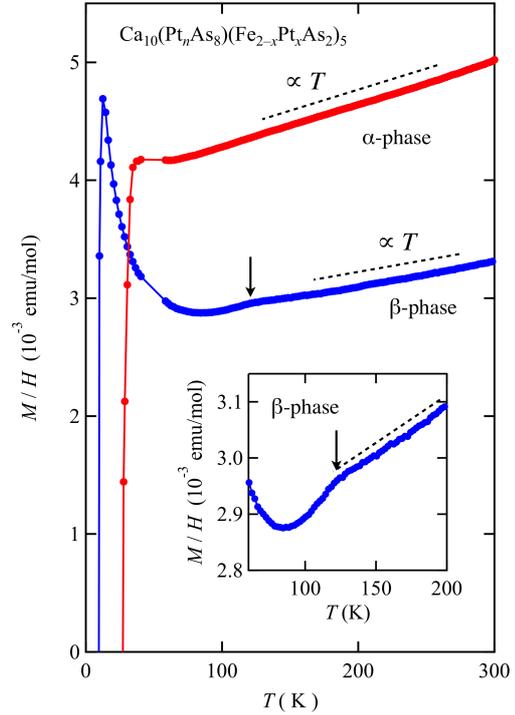}
\caption{
\label{}
Temperature dependence of magnetization divided by $H$, $M/H$, in a magnetic field of 1 T for a powder sample of Ca$_{10}$(Pt$_n$As$_8$)(Fe$_{2-x}$Pt$_x$As$_2$)$_5$ with $n$ = 4 ($\alpha$-phase) and $n$ = 3 ($\beta$-phase). The broken lines are guides for eyes. The allow indicates the temperature at magnetic anomaly.
}
\end{center}
\end{figure}

The temperature dependence of magnetic susceptibility, $M/H$, is shown in Fig.~6 for the $\alpha$- and $\beta$-phases. Susceptibility at high temperatures is characterized by the $T$-linear behavior, as indicated by the broken lines. Such $T$-linear behavior is unusual; however, it is widely observed in the normal state of iron-based superconductors \cite{0953-8984-22-20-203203}. Zhang {\it et al.} have shown theoretically that the $T$-linear behavior originates from short-range antiferromagnetic fluctuations \cite{0295-5075-86-3-37006}. 
The observed $T$-linear dependence indicates the existence of magnetic fluctuations in the present compounds. 

For the $\beta$-phase, Ca$_{10}$(Pt$_3$As$_8$)(Fe$_{2-x}$Pt$_x$As$_2$)$_5$, we observed an anomaly in susceptibility at approximately 120 K, as indicated by the arrow in Fig. 6. At the same temperature, the temperature coefficient of electrical resistivity changes from metallic behavior to a semiconducting one \cite{JPSJ.80.093704}. This behavior is analogous to those reported in the under-doped Ba(Fe$_{1-x}$Co$_x$)$_2$As$_2$ with $x$ = 0.05 \cite{JPSJ.78.123702}. The Hall measurements support that the specimen is in the under-doped regime \cite{JPSJ.80.093704}. 
We speculate that antiferromagnetic ordering sets in at approximately 120 K, and superconductivity at 14 K may coexist with antiferromagnetic ordering for the $\beta$-phase, Ca$_{10}$(Pt$_3$As$_8$)(Fe$_{2-x}$Pt$_x$As$_2$)$_5$;  further investigation is required to support our hypothesis.

\section{Role of Pt substitution}

Co-doped CaFe$_2$As$_2$ exhibits a maximum superconducting transition temperature $T_c$ = 20 K near the critical concentration of Co, 6\% ($x$ $=$ 0.06), at which the AFM ordering is completely suppressed in Ca(Fe$_{1-x}$Co$_x$)$_2$As$_2$ \cite{Harnagea}. Ni-doped CaFe$_2$As$_2$ exhibits similar behavior, while the critical concentration of Ni, at which the AFM phase is suppressed and superconductivity appears, is almost half of that for Co-doped CaFe$_2$As$_2$, i.e. 3\% ($x$ $=$ 0.03) in Ca(Fe$_{1-x}$Ni$_x$)$_2$As$_2$ \cite{Kumar}. 
This leads to a naive understanding that the dependence of $T_N$ and $T_c$ on doping level $x$ can be interpreted in terms of the difference in the number of valence electrons between the doped transition-metal element and Fe \cite{rf:Saha,rf:Ni}. 

Pt and Ni are isovalent elements. Thus, we may naively expect that Pt doping of approximately 3\% will be enough to suppress AFM ordering and to induce superconductivity in CaFe$_2$As$_2$ as well as in Ca$_{10}$(Pt$_n$As$_8$)(Fe$_{2-x}$Pt$_x$As$_2$)$_5$. 
In contradiction to this expectation, however, what we observed is a requirement of heavy Pt doping for superconductivity: 
A doping level of 8\% ($x$ $=$ 0.16) is not sufficient to suppress antiferromagnetic ordering in $\beta$-Ca$_{10}$(Pt$_3$As$_8$)(Fe$_{2-x}$Pt$_x$As$_2$)$_5$. 
A doping level of 18\% ($x$ $=$ 0.36) is necessary to induce superconductivity at $T_c$ = 38 K in $\alpha$-Ca$_{10}$(Pt$_4$As$_8$)(Fe$_{2-x}$Pt$_x$As$_2$)$_5$. 
Such ineffectiveness of Pt can be also seen in CaFe$_2$As$_2$ \cite{Kudo}: The AFM phase persists until the Pt doping level reaches its solubility limit at 8\% ($x$ $=$ 0.08) in Ca(Fe$_{1-x}$Pt$_x$)$_2$As$_2$. Superconductivity is absent in Ca(Fe$_{1-x}$Pt$_x$)$_2$As$_2$ up to $x$ = 0.08. 
It is interesting to note here that an attempt to dope Pt beyond the solubility limit at $x$ = 0.08 yields $\beta$-Ca$_{10}$(Pt$_3$As$_8$)(Fe$_{2-x}$Pt$_x$As$_2$)$_5$ ($x/2$ = 0.08) together with Ca(Fe$_{1-x}$Pt$_x$)$_2$As$_2$ ($x$ = 0.08). The former exhibits superconductivity at 13 K \cite{JPSJ.80.093704}, while the latter is not \cite{Kudo}, although the Pt content of the Fe site is almost the same (8\%). 
These observations will give us an unique opportunity to elucidate the role of chemical doping in the occurrence of superconductivity in iron-based materials.

\section{Conclusions}

In this paper, we provided an overview of the crystal structures and physical properties of the newly discovered superconductors, quaternary iron-platinum-arsenides  Ca$_{10}$(Pt$_4$As$_8$)(Fe$_{2-x}$Pt$_x$As$_2$)$_5$ ($\alpha$-phase) and Ca$_{10}$(Pt$_3$As$_8$)(Fe$_{2-x}$Pt$_x$As$_2$)$_5$ ($\beta$-phase). 
The compounds can be characterized by the platinum-arsenide layers composed of As$_2$ dimers, Pt$_n$As$_8$, which alternate with superconducting Fe$_2$As$_2$ layers. 
The As-Fe-As bond angle of the Fe$_2$As$_2$ layers is close to the ideal value for the $\alpha$-phase. This, together with the appropriate electron doping, makes the system high-$T_c$ up to 38 K.
We observed upper critical field $H_{c2}$ comparable with those reported in BaFe$_2$As$_2$, negative pressure coefficient of $T_c$, and normal-state magnetic susceptibility with characteristic $T$-linear behavior, indicative of magnetic fluctuations.
The next step is to control the Pt content in the Fe$_{2-x}$Pt$_x$As$_2$ layers in order to reveal the electronic phase diagram.
Another challenge is to modify the Pt$_n$As$_8$ spacer layers to make them more insulating or metallic in terms of conductivity to see whether the superconducting transition temperature can be enhanced to higher than 38 K.  



\section{Acknowledgments}

Part of this work was performed at the Advanced Science Research Center, Okayama University. 
This work was partially supported by KAKENHI from JSPS and MEXT, Japan.






\end{document}